\documentclass[aps,prb,superscriptaddress,reprint]{revtex4-1}
\usepackage{amsmath,amssymb} % mathematics
\usepackage{amsmath} % mathematics
\usepackage{amsfonts} % mathematics
\usepackage{bm} % bold in math-mode
\usepackage[colorlinks=true,citecolor=blue,linkcolor=blue,filecolor=blue,urlcolor=blue]{hyperref} % hyperlink with color

\usepackage[pdftex]{graphicx} % figure
\usepackage{braket} % braket
\usepackage{mathtools}

\usepackage{xcolor}
\usepackage{ulem}

\begin{document}
%%%%%%%%%%%%%%%%%%%%%%%%%%%%%%%%%%%%%%%%%%%%%%%%%%%%%%%%%%%%
%%%%%%%%%%%%%%% title and author information %%%%%%%%%%%%%%%
%%%%%%%%%%%%%%%%%%%%%%%%%%%%%%%%%%%%%%%%%%%%%%%%%%%%%%%%%%%%
% Title
\title{Specific heat and susceptibility of $S=1/2$ antiferromagnets on square, triangular, and kagome lattices}
% Author information
\author{Chisa Hotta}
\email{chisa@phys.c.u-tokyo.ac.jp}
\affiliation{Department of Basic Science, University of Tokyo, Meguro-ku, Tokyo 153-8902, Japan}
\date{\today}

%%%%%%%%%%%%%%%%%%%%%%%%%%%%%%%%%%%%%%%%
%%%%%%%%%%%%%%% abstract %%%%%%%%%%%%%%%
%%%%%%%%%%%%%%%%%%%%%%%%%%%%%%%%%%%%%%%%
\begin{abstract}
We study the temperature dependence of the thermodynamic properties of spin-1/2 antiferromagnets 
on two-dimensional lattices. 
Our analysis employs the sine-square deformation (SSD), 
in which a real-space envelope function is applied to the Hamiltonian so that 
the local energy scale is smoothly reduced to zero at the system boundaries. 
The quantum eigenstates of the SSD Hamiltonian exhibit bulk-like behavior near the system center, 
effectively mimicking the thermodynamic limit even in small finite-size calculations. 
Using these fictitious bulk states, we compute the energy density, 
specific heat, and magnetic susceptibility as functions of temperature. 
We find that both the triangular- and kagome-lattice antiferromagnets show 
either a shoulder or a pronounced double-peak structure in the low-temperature specific heat, 
whereas the kagome case particularly shows a strong enhancement of magnetic susceptibility 
down to the lowest temperature range. 
These direct comparisons, together with the square-lattice and one-dimensional cases, reveal that 
although both frustrated systems retain a substantial amount of entropy, 
the low-energy excitations below $\sim 0.5J$ of the kagome lattice are 
predominantly governed by the magnetic excited states, whereas not much for the triangular lattice. 
\end{abstract}

\maketitle
\section{Introduction}

Specific heat and static magnetic susceptibility are the very basic thermodynamic quantities 
for characterizing the nature of materials, 
and evaluating their accurate behavior down to the lowest temperatures 
has been a long-sought goal in theories. 
For instance, the temperature dependence of magnetic susceptibility 
above its peak provides direct information on the underlying microscopic interactions, 
while the low-temperature onset of the specific heat, 
namely whether it follows a power law or an exponential form, 
offers insight into the nature of elementary excitations. 
Indeed, for the kagome antiferromagnet, there had been long-standing discussions on the 
low energy properties\cite{lauchli2019}, 
which may have recently converged toward a gapless quantum-spin-liquid picture \cite{yin-chen2017,xiang2017}.
A wide variety of kagome materials, including herbertsmithite \cite{mendels2010,mendels2022}, 
volborthite \cite{watanabe2016}, kapellasite \cite{doki2018}, 
and more recently, the YCOB and In-kapellasite families exhibiting magnetization plateaus 
\cite{zheng2025,kato2024}, continue to stimulate active research. 
Similarly, triangular-lattice antiferromagnets such as 
Li$_2$AMo$_3$O$_8$ \cite{iida2019} and $A$YbSe$_{2}$ ($A$= Na, K)\cite{scheie2024}, 
which may lie close to a spin-liquid regime \cite{kaneko2014}, have attracted considerable attention. 
If theoretical approaches could reliably provide thermodynamic quantities 
suitable for quantitative comparison with experiments \cite{mendels2022}, 
they would offer important clues for elucidating the intriguing properties 
of these materials and of the underlying lattice models. 
\par
From early on, the standard approach for evaluating thermodynamic quantities 
was the high-temperature series expansion (HTE)\cite{elstner1994,oitmaa1996}, 
describing them as power series in the inverse temperature, $\beta=1/k_BT$. 
Typical expansions reach $\beta^{20}$ for one-dimensional(1D) systems \cite{elstner2000}
 and up to $\beta^{17}-\beta^{20}$ for two-dimensional quantum magnets \cite{pierre2024}, 
which roughly account for the spatial range over correlation of 10 lattice distance. 
As a result, the reliability of the HTE is limited down to temperatures of the order 
of the exchange interaction, $k_BT \gtrsim J$\cite{pierre2024}.
To access lower temperatures, one usually applies Pad\'e approximants, 
but these eventually become strongly parameter-dependent and may deviate significantly 
from the true thermodynamic behavior. 
To overcome this limitation, Bernu and Misguich proposed an interpolation scheme 
\cite{bernu2000,schmidt2017} that incorporates physical insight 
into the low-temperature form of the specific heat. 
By combining HTE data with sum rules and assumed low-$T$ functional behavior, 
they constructed interpolated curves that smoothly connect the high-temperature expansion 
with the ground-state limit.
The numerical linked-cluster (NLC) method pushed forward the conventional HTE 
by systematically including cluster contributions evaluated by exact diagonalization.
This approach reaches relatively lower temperatures when 
applied to several frustrated lattices \cite{rigol2006,rigol2007}. 
\par
Regarding numerical approaches for quantum many-body systems on finite clusters, 
the most fundamental methods are finite-temperature exact diagonalization (ED) \cite{jakelic1994} 
and its extensions \cite{hans1993,hams2000,iitaka2003}, 
including the family of thermal pure quantum (TPQ) state techniques 
\cite{sugiura2012,sugiura2013}, as well as transfer-matrix approaches \cite{imada1986}.
Because these methods explicitly treat the full Hilbert space, whose dimension grows 
exponentially with the system size $N$, 
they are practically limited to typically $N\lesssim 30$ and maximally $\sim 40$. 
This makes it difficult to resolve the delicate low-temperature behavior 
below the energy scale of the exchange interaction. 
To date, the only unbiased numerical technique capable of accessing the entire temperature 
range is quantum Monte Carlo (QMC) \cite{sandvik1999}.
However, except for special cases such as the square-lattice 
Heisenberg antiferromagnet \cite{ding1991}, most of the systems of current interest suffer 
from severe sign problems, placing them outside the reach of QMC despite its potential accuracy. 
\par
One dimensional (1D) system is rather an exception; several canonical integrable models admit exact finite-temperature solutions, 
e.g. the quasiparticle spin-1/2 Heisenberg and XXZ models can be solved exactly using the thermodynamic Bethe ansatz \cite{klumper1993,klumper1998}, and we use these results as benchmarks in the present work. 
Finite-temperature density-matrix renormalization group (DMRG) methods are also well 
established\cite{shibata1997,feiguin2005}. 
More recently, tensor-network approaches based on matrix product states (MPS) 
and matrix product operators (MPO) \cite{verstraete2004} have undergone significant development. 
These include TPQ-MPS \cite{iwaki2021,iwaki2022}, 
minimally entangled typical thermal states (METTS) \cite{white2009,stoudenmire2010}, 
and the exponential thermal renormalization group (XTRG) \cite{chen2018,li2019}.
Related numerical techniques can sometimes be extended beyond strictly 1D systems 
to some limited numbers of models like square\cite{poilblanc2021} and Kitaev magnets\cite{gohlke2023}, 
as well as Shastry-Sutherland model\cite{wietek2019,jimnez2021}. 
\par
These numerical approaches are all based on the idea of reducing the effective basis dimension, 
guided by the intuition that only a tiny fraction of the full Hilbert space contributes significantly 
to the ground state or even to finite-temperature pure or purified states \cite{barthel2017}.
However, any approximate reduction of the basis inevitably carries the risk of either missing 
important components or overemphasizing particular basis states.
\par
Finite-size effects represent one major source of unavoidable missing information, 
regardless of the quality of approximation. 
In translationally invariant systems, the allowed momenta $k$ become discrete, 
resulting in an energy discretization that follows the well-known $1/N$ 
scaling in 1D\cite{fisher1972,cardy1986}.
In 2D, the shape or aspect ratio of the finite-size cluster matters\cite{sandvik2012,nishimoto2013}, 
and requires size scaling over the lengthscale of $\gtrsim 20$. 
\par
In this paper, we apply the sine-square deformation (SSD) technique \cite{gendiar2009} 
to maximally reduce finite-size effects and provide a shortcut 
to directly accessing nearly bulk thermodynamic properties \cite{gendiar2011,hotta2012,hotta2013}. 
SSD offers a general framework that can be combined with any numerical solver 
that selects an optimized minimal basis for representing finite-temperature pure states,  
and enables direct evaluation of physical quantities without the need for conventional size scaling.
\par
We benchmark the method on the $S=1/2$ antiferromagnetic (AF) Heisenberg models in a 1D chain and on a square lattice 
to demonstrate its efficiency, and then apply it to the triangular and kagome lattices. 
In what follows, we first overview the role of SSD and present the main results 
in \S.\ref{sec:overview}, with their full information provided in comparisons with earlier results 
from other techniques in \S.\ref{sec:magnets}. 
In \S. \ref{sec:formulation} we explain the technical detail the SSD framework and explain how it is incorporated 
into methods that construct pure quantum states as well as those accessing the Gibbs ensemble.

%*%*%*%*%*%*%*%*%*% fig1 %*%*%*%*%*%*%*%*%*%
\begin{figure}[tb]
\centering
\includegraphics[width=0.45\textwidth]{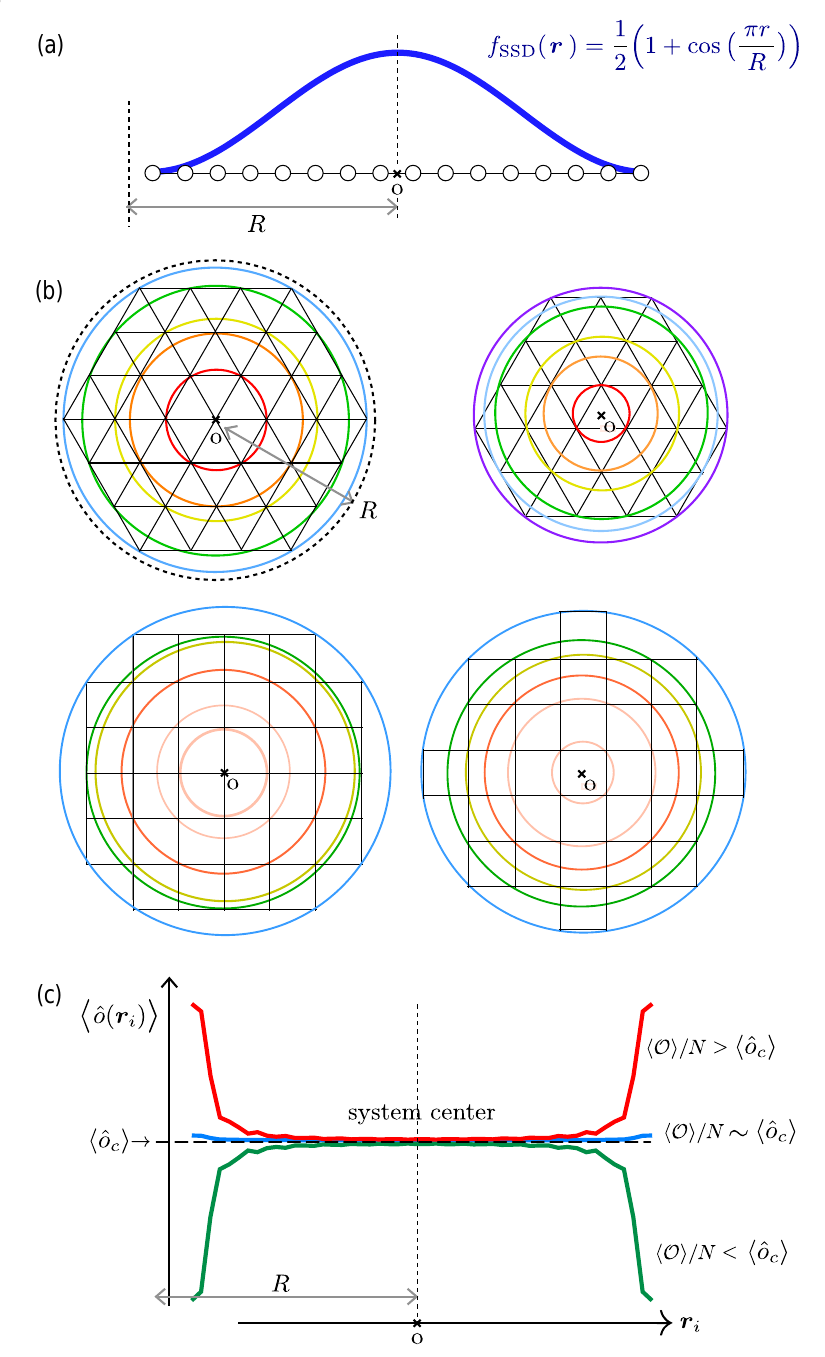}
\caption{
(a) Sine-square deformation (SSD) function, $f_{\rm SSD}$, defined along the 1D cut of the cluster 
which crosses the system center $o$. 
(b) Examples of polar SSD on 2D square and triangular lattices. 
Left/right-hand side ones are the site/bond-centered SSD. 
(c) Schematic illustration of how the physical quantities $\langle \hat o(\bm r_i)\rangle $ 
distributes in space. The value at the center $r_i\sim 0$ is insensitive to the 
total value of $\langle {\cal O} \rangle= \sum_i \hat o(\bm r_i)$, 
which we want to extract as $\langle \hat o_c\rangle$. 
}
\label{f1}
\end{figure}
%*%*%*%*%*%*%*%*%*%%*%*%*%*%*%*%*%*%*%%*%*%*%

%*%*%*%*%*%*%*%*%*% fig8 %*%*%*%*%*%*%*%*%*%
\begin{figure*}[tb]
\centering
\includegraphics[width=0.9\textwidth]{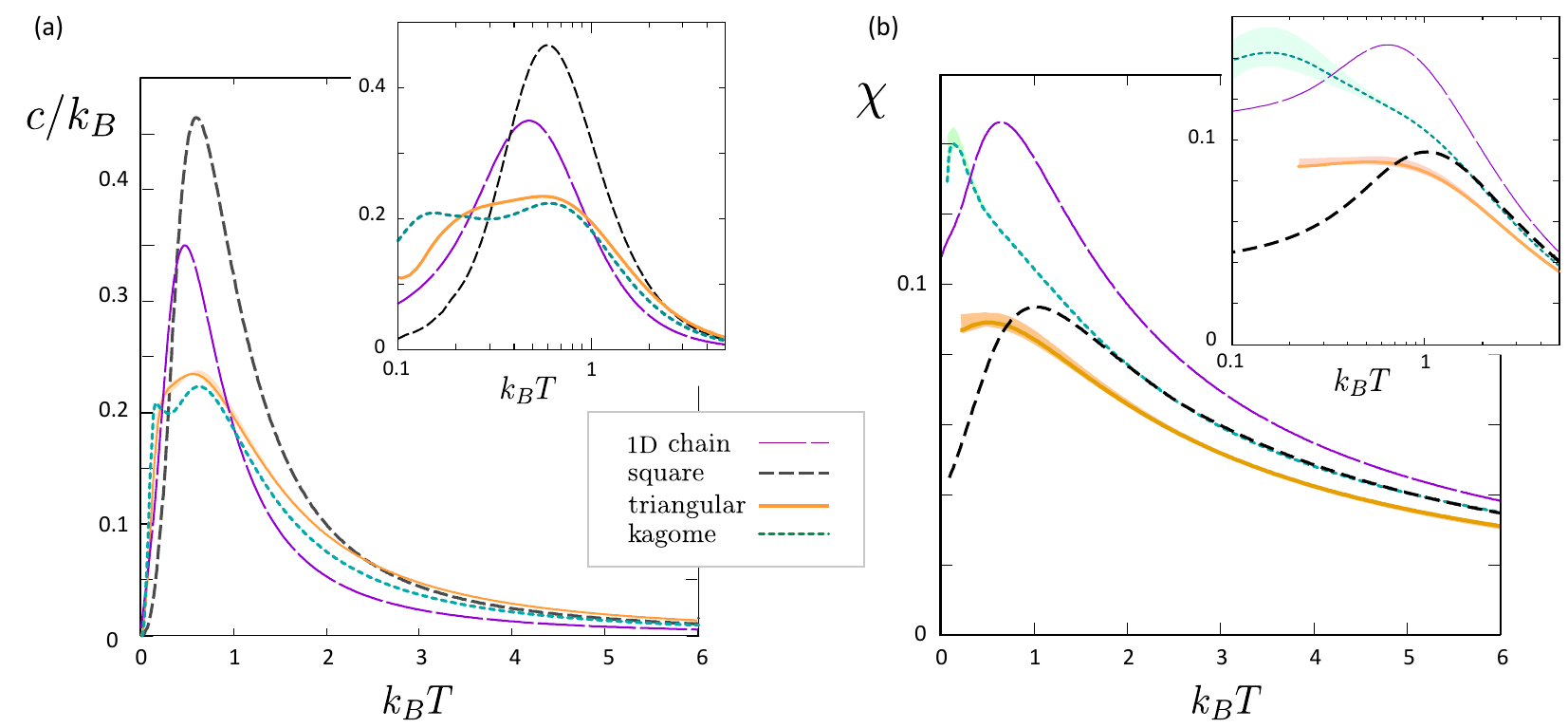}
\caption{Comparison of the (a) Specific heat and (b) susceptibility of 
the $S=1/2$ Heisenberg model on a 1D chain, square, triangular, and kagome lattices, 
obtained by TPQ method combined with SSD on a cluster of $N\le 27$. 
More details are found in Figs.\ref{f2}-\ref{f4}. 
Inset shows the same plot in the logarithmic scale of $k_BT$. 
}
\label{f8}
\end{figure*}
%*%*%*%*%*%*%*%*%*%%*%*%*%*%*%*%*%*%*%%*%*%*%iables are derived from $\Omega(\mu,T)$. 

%*%*%*%*%*%*%*%*%*%%*%*%*%*%*%*%*%*%*%%*%*%*%*%*%*%*%*%*%*%*%*%*%%*%*%*%*%*%*%*%*%*%%*%*%*%
%*%*%*%*%*%*%*%*%*%%*%*%*%*%*%*%*%*%*%%*%*%*%*%*%*%*%*%*%*%*%*%*%%*%*%*%*%*%*%*%*%*%%*%*%*%
\section{Overview of sine-square deformation}
\label{sec:overview}
\subsection{sine-square deformation}
We first outline our framework, and show the key results that we obtained using it. 
We consider a finite size 1D chain or 2D cluster shown in Figs.~\ref{f1}(a) and \ref{f1}(b) and 
define an SSD function given as 
\begin{equation}
 f_{\rm SSD} (\bm r_i)= \frac{1}{2}\Big(1+\cos \big(\frac{\pi r_i}{R}\big)\Big), 
\label{fssd}
\end{equation}
where we take the origin of $\bm r_i$ as the center of the open boundary finite-size cluster, 
with $R$ chosen to be the distance from the origin to the farthest edges of the cluster or slightly larger, 
e.g. larger by half-lattice spacing. 
The SSD function takes the maximum at the system center, and smoothly scales down toward the edges. 
We use this function as an envelope function of the Hamiltonian, 
namely multiplying this function to each local operator term of the Hamiltonian 
defined at $\bm r_i$. 
By solving this SSD Hamiltonian, physical quantities can be obtained in a manner 
equivalent to those of the Hamiltonian before the deformation. 
Since the SSD Hamiltonian becomes site dependent on top of the open boundary condition (OBC), 
the wave number $k$ is no longer a quantum number. 
Figure~\ref{f1}(c) illustrates the typical spatial profile of the density of extensive physical quantities, 
such as energy, particle number, and magnetization, measured at $\bm r_i$. 
Although these quantities are conserved globally, their spatial distribution is not homogeneous. 
\par
It was shown that the particle or magnetization density near the system 
center does not change significantly with the total particle number or magnetization (see Fig.~\ref{f1}(c)). 
These central values mimic those in the thermodynamic limit\cite{hotta2012,hotta2013}. 
The implication of this deformation was clarified using free fermions \cite{hotta2013}: 
different $k$ eigenstates of the original Hamiltonian mix via scattering induced by $f_{\rm SSD}$ 
and form a wave packet. The eigenstates of the deformed Hamiltonian can then be described 
using a small number of such wave-packet states. 
The location of the wave packet in real space roughly corresponds to the energy scale set by 
$f_{\rm SSD}(\bm r_i)$; thus, those near the system center represent the energy scale of the 
original Hamiltonian, where we intend to extract the intrinsic physical quantities. 
Because the wave packets are semi-localized, they do not suffer from strong finite-size effects 
characteristic of plane waves. As a result, the obtained physical quantities are significantly 
less affected by system size.
\par
One may then ask whether there is anything special about $f_{\rm SSD}$, 
and whether it can be replaced by other windowing functions. 
In Ref.~[\onlinecite{hotta2013}], we tested several different functional forms and found that 
other functions tend to delay numerical convergence and yield less accurate results. 
The underlying reason lies in how the wave packets are formed: $f_{\rm SSD}$ induces mixing of 
$k$ states only within nearby momenta. 
This is because its Fourier transform has a minimum wave number of order $(2R)^{-1}$, 
which corresponds to the smallest meaningful momentum scale set by the inverse system size. 
\par
\subsection{temperature dependent thermodynamic quantities}
We first summarize briefly the main results of this paper in Fig.~\ref{f8}, 
which shows the specific heat and uniform susceptibility 
of the 1D, square-, triangular-, and kagome-lattice $S=1/2$ Heisenberg antiferromagnets. 
These results are obtained by TPQ calculations for clusters up to 27 sites combined with SSD, 
using the full Hilbert space, whose computational cost is comparable to that of ground-state ED. 
This figure highlights the distinct thermodynamic features arising from different lattice geometries. 
\par
A first notable feature of frustrated magnetism is that the specific heat is strongly 
suppressed down to very low temperatures, showing an extended and relatively low peak. 
For the susceptibility, the kagome lattice almost coincides with the square lattice for 
$T/J \gtrsim 2$, presumably because the coordination numbers of the two lattices are the same 
and only short-range correlations over one lattice spacing are relevant in this temperature regime. 
The triangular and square lattices exhibit susceptibility peaks of comparable height, 
consistent with the fact that both lattices host magnetic long-range order at zero temperature 
\cite{reger1988,huse1988,bernu1992,jolicoeur1989,bernu1994,white2007}. 
In contrast, the kagome lattice shows a substantial enhancement of $\chi$, though not as dramatic as in 1D, 
indicating that magnetic fluctuations persist down to low temperatures. 
\par
The details of these results, together with comparisons to previous studies using other methods, 
are presented in \S\ref{sec:2d-magnets}.
\par
Although in the present work we restrict ourselves to numerical solvers that treat 
the full Hilbert space, namely, full ED and TPQ calculations up to $N=27$, 
the resulting thermodynamic quantities already capture the intrinsic features 
of the thermodynamic limit, in some cases even quantitatively. 
We emphasize that SSD is compatible with better solvers capable of treating significantly larger system sizes
ideally $N \gtrsim 100$, in which case 
one can expect an almost perfect reproduction of the true thermodynamic-limit behavior, 
as previously demonstrated for the ground state\cite{nishimoto2013}. 
So far, however, finite-temperature solvers that could be combined with SSD remain limited. 
One is the aforementioned XTRG\cite{li2019,li2020,li2020-2}, 
which has been applied to the square, triangular, and Kitaev honeycomb models with system sizes of $N=50-70$. 
Another is the TPQ method, which we briefly introduce below. 
For the Kitaev honeycomb model, the TPQ approach with a cylindrical geometry for 2D systems 
has successfully reproduced the characteristic double-peak structure of the specific heat for
$N=64$ \cite{gohlke2023}. 
Once they are combined with SSD, one can expect results obtained with the same cluster sizes 
to approach the bulk limit much more closely.

%*%*%*%*%*%*%*%*%*%%*%*%*%*%*%*%*%*%*%%*%*%*%*%*%*%*%*%*%*%*%*%*%%*%*%*%*%*%*%*%*%*%%*%*%*%
%*%*%*%*%*%*%*%*%*%%*%*%*%*%*%*%*%*%*%%*%*%*%*%*%*%*%*%*%*%*%*%*%%*%*%*%*%*%*%*%*%*%%*%*%*%
\section{general framework}
\label{sec:formulation}
In this section, we explain how to obtain the thermodynamic quantities we briefly showed in the previous section 
using the SSD Hamiltonian. 

\subsection{Deriving physical properties using SSD}
\label{sec:ssdgs}
We consider a general form of the lattice Hamiltonian with short-range interactions, 
given as a sum of local operators, 
\begin{align}
&{\cal H}=\sum_{i} \hat h(\bm r_i) - {\cal F} \sum_i \hat o(\bm r_i), 
\end{align}
where $\hat h(\bm r_i)$ is defined either on each site or on the center of the bond located at $\bm r_i$. 
We introduce a local operator $\hat o(\bm r_i)$ coupled to the field ${\cal F}$. 
For an electronic system, ${\cal F}$ correspond to the chemical potential term $\mu$ coupled to 
the particle density operator $n(\bm r_i)$. 
For the magnetic system, ${\cal F}$ and $\hat o(\bm r_i)$ are read off as the magnetic field $h$ and 
the magnetization density $m(\bm r_i)$, respectively. 
\par
In general, one often considers the case where $\langle {\cal O} \rangle =\sum_i \hat o(\bm r_i)$ 
is a conserved quantity fulfilling $[{\cal O}, {\cal H}]=0$, and accordingly, 
$\langle {\cal O}\rangle$ takes constant value in each block-diagonalized subspace. 
Indeed, the aforementioned particle number and total magnetization are conserved for the Hubbard or Heisenberg models. 
The ground state for a given ${\cal F}$ is determined for $\langle {\cal O}\rangle$
that minimizes $\langle {\cal H}\rangle$, and by extrapolating this ground state quantities to $N\rightarrow\infty$, 
we obtain $\langle {\cal O} \rangle/N$ as a response to ${\cal F}$. 
However, due to the limitation of size $N$, this finite-size scaling often gives undecisive results, 
particularly in 2D, as the results depend seriously on both the shape and the limited sizes of the clusters. 
The SSD gives a recipe to solve this long-standing issue, as we explain in the following. 
\par
The SSD multiplies each term by a smooth envelope function, $f_{\rm SSD}$, as
\begin{align}
&{\cal H}_{\rm SSD}= \sum_{\bm i}f_{\rm SSD} (\bm r_i) h(\bm r_i) -  {\cal F} \sum_i f_{\rm SSD} (\bm r_i) \hat o(\bm r_i) , 
\label{hssd}
\end{align} 
Suppose that we have the ground state of ${\cal H}_{\rm SSD}$, 
defined as $|\Phi_0 \rangle$, with energy $E_0$. 
Again, we would like to evaluate an extensive physical quantity, ${\cal O} =\sum_i \hat o(\bm r_i)$. 
As mentioned above, this type of quantity includes particle number, magnetization, and bond energy, etc. 
The expectation values in terms of the SSD ground state largely depend on ${\bm r}_i$, 
and we are interested in the one we obtain at the system center. 
This center value is expected to mimic $\langle o_i \rangle$ for a periodic boundary condition (PBC) in the thermodynamic limit 
within sufficient accuracy, as demonstrated in our series of works
\cite{hotta2012,hotta2013,nishimoto2013,hotta2018}. 
\par
We first calculate the expectation value, $\langle \hat o(\bm  r_i)\rangle_0=\langle \Psi_0|\hat o(\bm  r_i)| \Psi_0\rangle$, 
of the ground state all over the system. 
For the quantum many-body systems, $\langle \hat o(\bm  r_i)\rangle$ shows an oscillatory behavior as a function of 
distance $|\bm r_i|$ from the origin at the system center, where the oscillation amplitude is the smallest at $|\bm r_i|\sim 0$. 
The oscillation occurs due to the interplay of open boundary effect and strong correlation, which becomes smaller for larger system size\cite{shibata2011}. 
Ideally, we want to extract the value at the system center after smoothing out this oscillation. 
There are two ways to extract it. 
One is to take the weighted average over SSD function as 
\cite{kawano2022},  
\begin{align}
\langle \hat o_c\rangle_{\rm SSD} = \frac{\sum_i f_{\rm SSD}(\bm r_i) \langle \hat o(\bm  r_i)\rangle}
{\sum_i f_{\rm SSD} (\bm r_i)}. 
\label{eq:ssdfr0}
\end{align}
This equation is the zero-wave-number Fourier transformation weighted by $f_{\rm SSD}$, 
and automatically extracts the uniform component of the data 
that corresponds to the oscillatory-center value at $\bm r_i\sim 0$. 
The second way is to straightforwardly fit the data, by carefully choosing the 
cut off value of $|\bm r_i|$ to fall closely to the oscillatory period
(see Ref.[\onlinecite{hotta2012}] for details). 
For each given applied field, ${\cal F}$, we perform this analysis and extract $\langle \hat o_c\rangle_{\rm SSD}$. 
\par 
The application is straightforward. 
For example, for a quantum spin system, we may solve ${\cal H}_{\rm SSD}$ for a given magnetic field $h$, 
using DMRG, and from its ground state, evaluate the magnetization density $\langle m(\bm r_i)\rangle$ 
and extract its spatially uniform component using Eq.(\ref{eq:ssdfr0}) that 
is typically $\sim \langle m(\bm r_i=0)\rangle$. 
Plotting this value by varying $h$, one is able to obtain a smooth magnetization curve, 
which successfully figured out the existence of 1/9 plateau in the spin-1/2 kagome antiferromagnet
\cite{nishimoto2013}. 
Similarly, for a Hubbard model, obtaining a finite-$N$ ground state and by measuring the particle density 
at the system center, and varying a chemical potential $\mu$, one is able to recover a $\mu-n$ curve 
that shows a Mott transition. The finite charge gap is evaluated as the width of the plateau 
of this curve at half-filling\cite{hotta2012}. 
\par 
In finite lattices, the Hamiltonian allows for several boundary conditions, 
typically PBC, OBC, or twisted boundaries. 
These boundaries constitute an order-1 perturbation, and the effects of OBC 
edges decay only as Friedel oscillations proportional to the inverse distance. 
it is generally expected that physical observables should converge to the same thermodynamic values. 
From this viewpoint, one may ask whether SSD can be regarded as yet 
another type of boundary condition, effectively a smooth, system-wide twist applied over a single wavelength. 
This point of view allows us to interpret the local density matrix at the system center, 
$\rho_{c}={\rm Tr}_{i\ne c} |\Psi_0\rangle\langle \Phi_0|$, ($c$: center site index) 
as the one intrinsic to the original ${\cal H}$. 
The expectation value thus obtained, ${\rm Tr} (\rho_c \hat o_c)$, equivalent to Eq.(\ref{eq:ssdfr0}), 
indeed turned out to mimic the corresponding quantity at $N\rightarrow\infty$. 
%*%*%*%*%*%*%*	
\subsection{SSD for finite temperature properties using pure states}
\label{sec:ssdft}
Similar treatment is possible at finite temperature. 
In quantum many-body systems, the density operator of Hilbert space dimension $\Lambda$ is 
given by 
\begin{align}
\rho_{\rm G}&= e^{-\beta{\cal H}}/Z(\beta) \notag\\
&=\sum_{\alpha=1}^\Lambda \Big(e^{-\beta{\cal H}/2}|\alpha\rangle \langle \alpha| e^{-\beta{\cal H}/2} \Big)\Big/Z(\beta), \\
& Z(\beta)=\sum_{\alpha=1}^\Lambda  \langle \alpha| e^{-\beta{\cal H}}|\alpha\rangle, 
\end{align}
representing the Gibbs state as a mixture of an exponentially large number of pure states. 
Whereas, it is known that the number of summation is being reduced, depending on the temperature $\beta^{-1}$ and 
on the choice of a set of pure states that are summed over\cite{iwaki2022}. 
Namely, if we choose $\{|\alpha\rangle\}$ as a classical basis set, the Gibbs ensemble is required, 
whereas if we choose instead a highly entangled state, the number of summations to mix them is reduced by 
several orders of magnitude. 
The number of summations required scales with the degree of purity of the finite temperature state, 
and one is able to construct a thermal equilibrium state of an arbitrary purity ranging 
from a zero-purity Gibbs state to the purity-1 thermal pure quantum (TPQ) state\cite{iwaki2022}. 
\par
The well-established method to obtain a TPQ state\cite{sugiura2013} is to start from ideally a Haar random state 
$|\gamma^{(i)} \rangle=\sum_{j=1}^\Lambda c_j^{(i)} |\alpha\rangle$, 
where $c_j$ is the complex number chosen from a random distribution whose choice is indexed by 
the superscript $(i)$. 
By performing an imaginary-time evolution, we obtain {\it an unnormalized} TPQ state as 
\begin{align}
|\beta^{(i)} \rangle = e^{-\beta{\cal H}/2}|{\gamma^{(i)}} \rangle, 
\end{align}
and the resultant density matrix is represented by 
\begin{align}
\rho_{\rm TPQ}= |\beta^{(i)} \rangle \langle\beta^{(i)} |\big/\langle \beta^{(i)} |\beta^{(i)} \rangle. 
\end{align}
The physical quantity with a support ranging within less than half of the system size is known to 
be equally evaluated by these two density operators as, 
$\langle\,\hat o \rho_{\rm TPQ} \rangle \sim \langle \hat o\,\rho_{\rm G}\rangle $, 
within a random fluctuation suppressed exponentially by $N$\cite{sugiura2012,sugiura2013}. 
This random fluctuation is temperature (or free-energy) dependent, and is physically intrinsic\cite{iwaki2022}, 
but is different from a standard finite-size effect that appears also in the ground state. 
To suppress the random fluctuations that are particularly enhanced at $N\lesssim 20$, 
one needs to take more than $M\gtrsim 50$ samples by  
starting from different choices of $\{|\gamma^{(i)} \rangle\}_{i=1}^{M}$, and taking the average as 
\begin{align}
\langle o (\bm r_i) \rangle_\beta= \sum_{i=1}^M \langle\beta^{(i)} |\hat o(\bm r_i) |\beta^{(i)} \rangle \Big/ 
\sum_{i'=1}^M\langle \beta^{(i')} |\beta^{(i')} \rangle. 
\label{eq:randomav}
\end{align}
Here, we need to take the summation over $i=1,\cdots M$ 
independently between the numerator and the denominator to properly suppress the variance
(see Ref.\cite{iwaki2022} for proof). 
With this, one can practically get an accurate evaluation for that system size and temperature. 
\par
The purpose of combining the TPQ state and the SSD is to further get rid of the finite-size effect, 
which is straightforward. 
By simply replacing the ground state $|\Psi_0\rangle$ with the TPQ state, $|\beta^{(i)}\rangle$, 
we are able to evaluate the expectation value using Eq.(\ref{eq:randomav}) at that temperature 
from a set of data $\{ \langle o (\bm r_i)\rangle_\beta \}$. 
By substituting this set to Eq.(\ref{eq:ssdfr0}) we obtain 
$\langle \hat o_c\rangle_{\beta,{\rm SSD}}$ for a given ${\cal F}$ at temperature $\beta^{-1}$. 
This value is regarded as $\langle {\cal O}\rangle/N$ at $N\rightarrow\infty$ for an applied field 
${\cal F}$ and at finite temperature, $\beta^{-1}$.

%*%*%*%*%*%*%*
\subsection{Free fermionic system}
\label{sec:freefermion}
The above-mentioned treatment is applied to general quantum many-body systems, 
for which it is generally difficult to obtain a full set of eigenstates due to the Hilbert space 
dimension $\Lambda$ exponentially growing with $N$. 
However, for a noninteracting system, one is able to construct a full set of many-body eigenstates, 
which is useful to validate our framework. 
Here, before going into the next section, we demonstrate the case of free fermions as the simplest example. 
The SSD Hamiltonian is given by
\begin{align}
{\cal H}_{\rm SSD}
 =& \sum_{\langle i,j\rangle} f_{\rm SSD}(\bm r_{ij})\Big( -t\, c_i^\dagger c_j +{\rm H.c.} \Big)
 \!-\!\mu \sum_j f_{\rm SSD}(\bm r_j) n_j ,
\label{eq:ssd_freeh}
\end{align}
where $t=1$ is the transfer integral, $c_i/c_i^\dagger$ is the annihilation/creation operator of a fermion on site $i$, 
and $\bm r_{ij}/\bm r_j$ locate at the center of the bond/site. 
Diagonalizing the representation of ${\cal H}_{\rm SSD}$ 
in terms of the single-particle basis $\{c_j^\dagger|0\rangle\}_{j=1}^N$ 
we obtain the energy eigenvalue, $\epsilon_\ell$, $\ell=1,\cdots,N$, labeled in the ascending order 
and the corresponding single-particle eigenstates, $d_l^\dagger |0\rangle=\sum_j \varphi_l^*(j) c_j^\dagger|0\rangle $. 
The ground-state wave function for a given chemical potential $\mu$ is given by 
filling the Fermi sea up to the zero energy level, 
\begin{align}
|\Phi_\mu\rangle
 = \prod_{\epsilon_l \le 0} d_l^\dagger |0\rangle
  = \prod_{\epsilon_l \le 0}\Big( \sum_{j=1}^N \varphi_l^*(j)c_j^\dagger\Big)|0\rangle, 
\end{align}
and the particle density at site $j$ is evaluated as 
\begin{eqnarray}
\langle n_j \rangle = \langle\Phi_\mu| n_j |\Phi_\mu\rangle = \sum_{\epsilon_\ell \le 0} 
 |\varphi_\ell(j)|^2,
\label{nonint_nmu}
\end{eqnarray}
and the value $\langle n_j \rangle$ near the system center, denoted as $\langle n_c\rangle_{\rm SSD}$, 
approximates the particle density in the bulk limit for a given $\mu$. 
We obtain $\langle n_c\rangle_{\rm SSD}$ by either 
applying Eq.(\ref{eq:ssdfr0}) or exracting the value near the system center for large enough $N$ as 
\begin{equation}
\langle n_{j_c} \rangle \approx  \sum_{\epsilon_\ell \le 0}|\varphi_\ell(j_c)|^2, 
\end{equation}
with $j_c$ being the center-site index. 

%*%*%*%*%*%*%*%*%*% fig5 %*%*%*%*%*%*%*%*%*%
\begin{figure}[tb]
\centering
\includegraphics[width=0.45\textwidth]{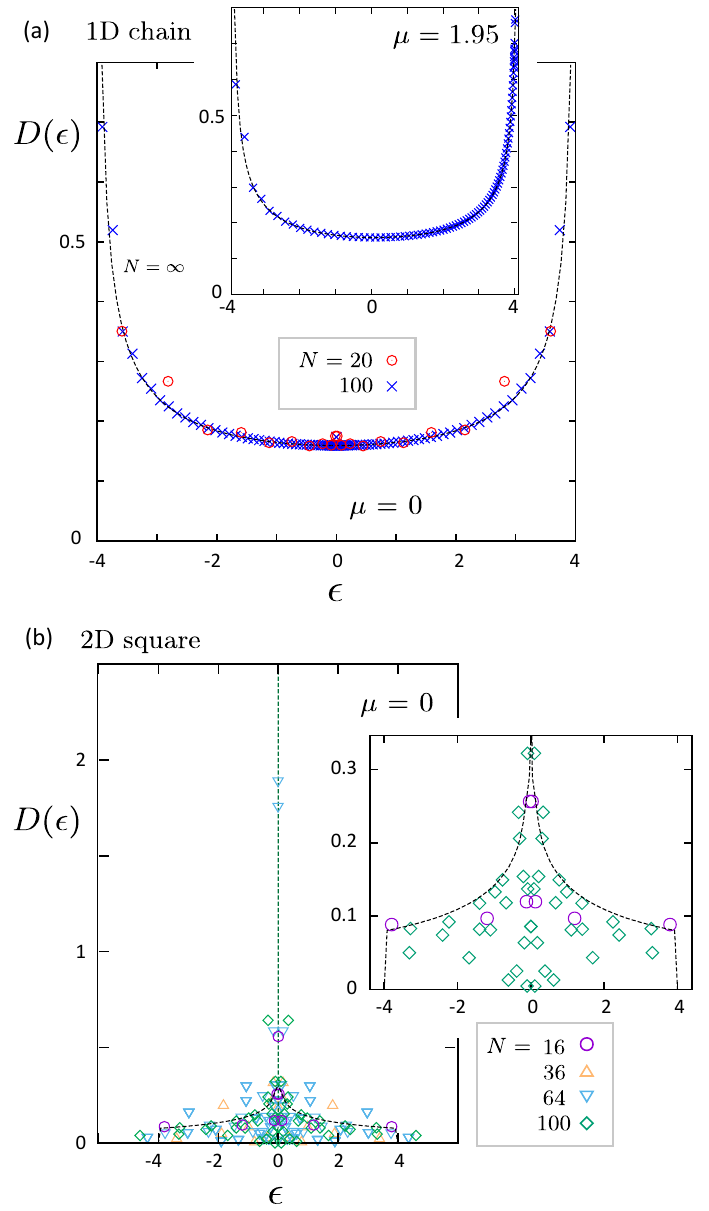}
\caption{
Density of states $D(\epsilon)$ (of single-particle energy eigenstate) of the noninteracting 
fermions obtained by ${\cal H}_{\rm SSD}$ 
in (a) 1D chain at $N=20, 100$ and (2) 2D square lattice with $N=16,36,64,100$, both at $\mu=0$. 
The inset of panel (a) is the case at $N=100$ and $\mu=1.95$. 
Broken lines are the exact DOS at $N=\infty$; 
for 1D we have $D(\epsilon)=\pi^{-1}\, d\arccos(-\epsilon/2)/d\epsilon$ 
}
\label{f5}
\end{figure}
%*%*%*%*%*%*%*%*%*%%*%*%*%*%*%*%*%*%*%%*%*%*%
%*%*%*%*%*%*%*%*%*% fig6 %*%*%*%*%*%*%*%*%*%
\begin{figure}[tb]
\centering
\includegraphics[width=0.45\textwidth]{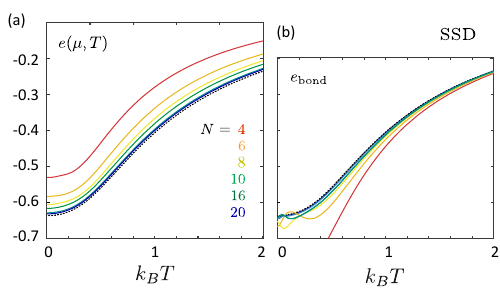}
\caption{Comparison of the evaluation of bond energy 
using the two formula, (a) $e(\mu,T)$ in Eq.(\ref{nonint_nmu}) 
and (b) $e_{\rm bond}$ in Eq.(\ref{eq:ebond}) 
for the same free fermion ${\cal H}_{\rm SSD}$ in 1D chain. 
We show several choices of $N=6,\cdots 20$, compared to 
$N=\infty$ in broken line. 
}
\label{f6}
\end{figure}
%*%*%*%*%*%*%*%*%*%%*%*%*%*%*%*%*%*%*%%*%*%*%
%*%*%*%*%*%*%*%*%*% fig7 %*%*%*%*%*%*%*%*%*%
\begin{figure*}[tb]
\centering
\includegraphics[width=1.0\textwidth]{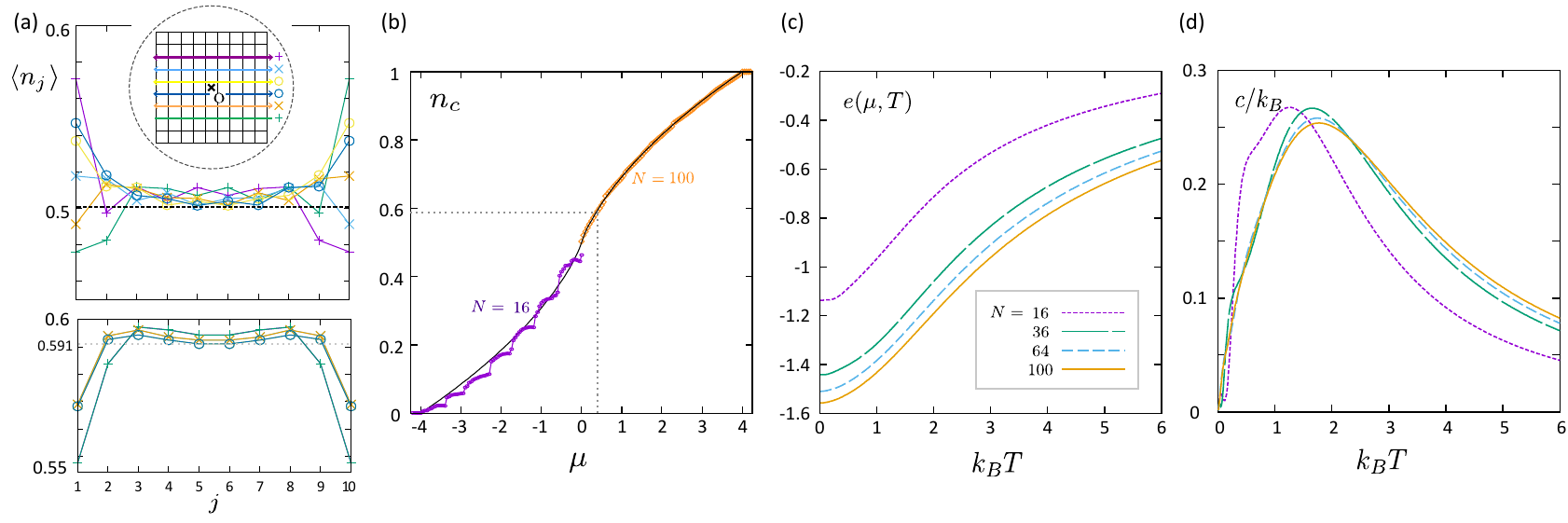}
\caption{Analysis of the square lattice free fermionic model using ${\cal H}_{\rm SSD}$. 
(a) $\langle n_j \rangle$ along the horizontal bonds on a $N=10\times 10$ square lattice 
at $\mu=0$ ($n_c=0.5$) and $\mu=0.4$ ($n_c=0.591$). 
(b) $\mu$-$n_c$ curve of the free fermion system on a square lattice. 
Data points are the center site value  $\langle n_c \rangle$ 
displayed for different lattice sizes, $N=4\times 4=16$ ($\mu\le 0$) and $N=10\times 10=100$ ($\mu\ge 0$). 
(c) Energy density $e_c$ and (b) specific heat $c/k_B=de/d(k_BT)$ as functions of temperature 
obtained using Eq.(\ref{eq:nonint-expct}) for $N=16,32,64,100$. 
}
\label{f7}
\end{figure*}
%*%*%*%*%*%*%*%*%*%%*%*%*%*%*%*%*%*%*%%*%*%*%
\par
We have shown previously\cite{hotta2013,hotta2018} that 
the single-particle density of states (DOS) of ${\cal H}_{\rm SSD}$ is given as 
its contribution from the $\ell$ th energy level as 
\begin{eqnarray}
D(\epsilon_\ell)
 \approx 
 \frac{
   |\varphi_l(j_c)|^2 
 + \big(|\varphi_{\ell+1}(j_c)|^2 + |\varphi_{\ell-1}(j_c)|^2\big)/2
 }{\epsilon_{\ell+1}-\epsilon_{\ell-1}}, 
\label{eq:dos}
\end{eqnarray}
where we average the neighboring three levels to smoothen the 
oscillations due to energy discretization. 
Figures~\ref{f5}(a) and \ref{f5}(b) show $D(\epsilon)$ obtained 
using Eq.(\ref{eq:dos}) for the 1D chain and square lattice with several different $N$. 
The agreement with the $N=\infty$ DOS (broken lines) of the corresponding lattices is almost perfect 
for 1D, and fairly well for the $N=10\times10=100$ square lattice. 
The remarkable feature is the dense distribution of the data points 
near $\epsilon \sim \mu$ (compare $\mu=0$ in the main panel and $\mu=1.95$ in the inset). 
The SSD nicely reproduces the bulk DOS, but is controlled in a way to 
make the energy discretization smaller at low energies near the chemical potential level. 
This fact supports the accuracy of SSD about describing the low-temperature physical quantities, 
and guarantees the reduction of the finite-size effect. 
\par
As the DOS is reproduced,  
the energy density $e(\mu,T)$ and physical quantities ${\hat o}_c$ 
at finite temperature are evaluated as 
\begin{align}
e(\mu,T)
 &= \sum_{\ell=1}^{N} (\epsilon_\ell +\mu)\, D(\epsilon_\ell)\, f(\epsilon_\ell),\\
 o_c (\mu,T)
 &= \sum_{\ell=1}^{N} \langle \hat o_c (\ell) \rangle_{\rm SSD} D(\epsilon_\ell)\, f(\epsilon_\ell),
\label{nonint_nmu}
\\
&f(\epsilon_\ell)=\frac{1}{e^{\beta \epsilon_\ell}+1}, 
\label{eq:nonint-expct}
\end{align}
where $f(\epsilon_\ell)$ is the Fermi distribution function  
with a notice that $\epsilon_\ell$ is measured from the chemical potential level $\mu$. 
As we are treating the grand canonical ensemble, 
when $o$ is taken as a particle density, $n_c$, 
it necessarily varies with $T$, unlike the natural assumption of fixed $n_c$ in the canonical system. 
\par
Here, we need to clarify that $e(\mu,T)$ in Eq.~(\ref{nonint_nmu}) is evaluated
using the one-particle energies $\epsilon_\ell$ defined over the entire system, 
and that $D(\epsilon_\ell)$ in Eq.~(\ref{eq:dos}) and in Fig.~\ref{f5} is a
function of $\epsilon_\ell$. 
At first sight, this seems to contradict the claim that the system center realizes the bulk limit. 
However, the $\ell$ th eigenstate of the SSD Hamiltonian is a wave packet localized around the region
where the deformed energy scale matches, namely, 
$|\varphi_\ell(j)|^2$ has its maximum weight at the site $j$ satisfying
$f_{\rm SSD}(\mathbf{r}_j)/t \sim \epsilon_\ell$ (see Ref.~\onlinecite{hotta2013}). 
Each $\epsilon_\ell$ corresponds to a well-defined local energy originating 
from a spatially localized eigenstate, and its DOS can be regarded as $D(\epsilon_\ell)$. 
To make this physical interpretation clearer, one may alternatively evaluate the 
bond energy directly as
\begin{align}
e_{\rm bond} 
&= \langle c_{j_c}^\dagger c_{j_c+1} + {\rm h.c.} \rangle \notag \\
&= \sum_{\ell=1}^{N} \left( 
    \varphi_{\ell}(j_c)^{*}\varphi_{\ell}(j_c+1)
    + {\rm h.c.}
    \right) f(\epsilon_\ell),
\label{eq:ebond}
\end{align}
where $D(\epsilon_\ell)$ is unnecessary, because the particle
density information is already encoded in the wave functions $\varphi_\ell$. 
As shown in Fig.~\ref{f6}, the two evaluation methods yield different
values for small system sizes $N$, with $e_{\rm bond}$ being overall more accurate. 
Nevertheless, both approaches converge rapidly toward the $N \to \infty$ value 
(indicated by the broken line). 
We confirmed that the evaluation of the bond energy using Eq.~(\ref{eq:int_eb}), 
which was formulated for interacting systems, agrees numerically exactly with 
$e_{\rm bond}$ when applied to the noninteracting Hamiltonian.  
\par
We finally demonstrate in Fig.~\ref{f7} the SSD analysis for free fermions on a square lattice.
Figure~\ref{f7}(a) shows the particle density $\langle n_j\rangle$ obtained using
Eq.~(\ref{nonint_nmu}) for an $N=10\times 10$ lattice.
At $\mu=0$, the density at the system center is close to $0.5$, and it becomes $0.591$ at $\mu=0.4$.
By extracting this center value for each $\mu$, we construct the $\mu$--$n_c$ curve shown in
Fig.~\ref{f7}(b).
The temperature dependence of $e(\mu,T)$ in Fig.~\ref{f7}(c) exhibits convergence with increasing $N$
in a manner similar to the 1D case shown in Fig.~\ref{f6}(a).
We further evaluate the specific heat from the temperature derivative,
$c/k_B = d e(\mu,T) / dT$, as displayed in Fig.~\ref{f7}(d).
The resulting behavior correctly captures the gapless nature of the system at low temperatures.

%*%*%*%*%*%*%*%*%*%*%*%*%*%*%*%*%*%*%*%*%*
\subsection{Interacting systems using Gibbs state or full ED}
We finally show another formulation to apply SSD to the interacting Hamiltonian, 
that makes use of the partition functions or a 
Gibbs ensemble, complementary to \S.\ref{sec:ssdft}.  
It corresponds to cases where we are able to handily obtain a full set of energy eigenstates 
for sufficiently small system sizes. 
The finite temperature Lanczos method (FTLM) \cite{jakelic1994}
and its extensions \cite{morita2020,hanebaum2014} that target the partition function 
also fit here by combining their algorithm with the following formula. 
\par
Let us consider a Hamiltonian of a particle system whose total particle number $N_p$ is conserved. 
For the quantum magnets, the total magnetization $S_z^{\rm tot}$ is often conserved, 
and the parallel discussion holds. 
One is able to separately diagonalize each sector labeled by a different $N_p$, 
whose dimension is $\Lambda(N_p)$; the diagonalization of ${\cal H}_{\rm SSD}$ yields 
the eigenvalues $E_{\ell}(N_p)$ in the ascending order, and many-body eigenstates 
$|\Psi_{\ell}(N_p)\rangle$ ($\ell=1,\dots,\Lambda(N_p)$). 
Let $N_p^{\rm (min)}$ denote the particle number minimizing $E_{1}(N_p)$ as, 
\begin{equation}
N_p^{\rm (min)} = {\mathop\mathrm{arg~min}\limits}_{N_p} E_{1}(N_p). 
\end{equation}
At $T=0$, the operator $\hat o(\bm r_i)$ acting on the center bond/site is evaluated as 
\begin{equation}
\langle \hat o(\bm r_i) \rangle_{\rm SSD}
 = \langle\Psi_{1}(N_p^{\rm (min)})|\hat o(\bm r_i)|\Psi_{1}(N_p^{\rm (min)})\rangle. 
\end{equation}
%%%%%
At finite temperature, the grand canonical partition function is given as 
\begin{align}
\Xi &= \sum_{N_p} \sum_{\ell=1}^{\Lambda(N_p)}
     e^{-\beta E_{\ell}(N_p)},
\label{int_nmu}
\end{align}
and the ensemble average of the operator $\hat o(\bm r_i)$ is 
\begin{align}
\langle \hat o(\bm r_i)\rangle_{\beta}&= \sum_{N_p,\ell} \langle \hat o(\bm r_i) \rangle_{\ell,N_p}  
 \frac{e^{-\beta E_{\ell}(N_p)}}{\Xi}, \notag\\ 
&  \langle \hat o(\bm r_i) \rangle_{\ell,N_p}  
 =\langle\Psi_{\ell}(N_p)| \hat o(\bm r_i) |\Psi_{\ell}(N_p)\rangle.
\label{eq:oc}
\end{align}
Substituting $\langle \hat o(\bm r_i)\rangle_{\beta}$ to Eq.(\ref{eq:ssdfr0}), 
we obtain $\langle o_c\rangle_{\beta,{\rm SSD}}$ for a given $\mu$ and $\beta=(k_BT)^{-1}$. 
\par
For example, for the tight-binding models, the bond energy is evaluated for a given $\mu$ as
\begin{align}
e_{\rm bond}(\mu) &= \sum_{N_p,\ell} \langle \hat e_{c} \rangle_{\ell,N_p}  
 \frac{e^{\beta E_{\ell}(N_p)}}{\Xi},
\\
&  \langle \hat e_{c}(\mu) \rangle_{\ell,N_p} 
 =  \langle\Psi_{\ell}(N_p)|
   c_{j_c}^\dagger c_{j_c+1}+{\rm h.c.}
 |\Psi_{\ell}(N_p)\rangle. 
\label{eq:int_eb}
\end{align}
If we apply Eq.(\ref{eq:int_eb}) to the noninteracting Hamiltonian Eq.(\ref{eq:ssd_freeh}) 
via the many-body formulation, 
we obtain exactly the same results as Fig.~\ref{f6}(b). 
\par
Once we obtain $\langle \hat n_{c}(\mu) \rangle_{\beta,{\rm SSD}}$ from Eq.(\ref{eq:oc}), 
one can read this off as a continuous function of $\mu$ 
and the thermodynamic potential follows 
\begin{align}
& \frac{\partial\Omega}{\partial\mu} = - N\langle \hat n_{c}(\mu) \rangle_{\beta,{\rm SSD}} 
\notag\\
& \quad\Rightarrow\quad
\Omega(\mu,\beta)= -N \int_{-\infty}^\mu \langle \hat n_{c} (\mu')\rangle_{\beta,{\rm SSD}}   \, d\mu'. 
\end{align}
Accordingly, the other thermodynamic variables are to be obtained once we 
are able to express $\Omega(\mu,\beta)$ as numerically smooth functions of $\mu$ and $k_BT=\beta^{-1}$.

%*%*%*%*%*%*%*%*%*%%*%*%*%*%*%*%*%*%*%%*%*%*%*%*%*%*%*%*%*%*%*%*%%*%*%*%*%*%*%*%*%*%%*%*%*%
%*%*%*%*%*%*%*%*%*%%*%*%*%*%*%*%*%*%*%%*%*%*%*%*%*%*%*%*%*%*%*%*%%*%*%*%*%*%*%*%*%*%%*%*%*%
%*%*%*%*%*%*%*%*%*%*%*%*%*%*%*%*%*%*%*%*%*%*%*%*%*%*%*%*
%*%*%*%*%*%*%*%*%*%*%*%*%*%*%*%*%*%*%*%*%*%*%*%*%*%*%*%*
\section{Specific heat and susceptibility of quantum antiferromagnets}
\label{sec:magnets}
We now present the temperature-dependent specific heat and susceptibility 
of the quantum $S=1/2$ antiferromagnetic Heisenberg model, 
\begin{equation}
{\cal H}=\sum_{\langle i,j\rangle} J \bm S_i \cdot \bm S_j - h\sum_{j} S_i^z, 
\end{equation}
defined on representative low-dimensional lattices of finite size $N$. 
Here, $S_i$ denotes the spin-$1/2$ operator and $h$ is an applied magnetic field, 
and the summation runs over all nearest-neighbor pairs $\langle i,j\rangle$. 

%*%*%*%*%*%*%*%*%*%%*%*%*%*%*%*%*%*%*%%*%*%*%
%*%*%*%*%*%*%*%*%*% fig2 %*%*%*%*%*%*%*%*%*%
\begin{figure*}[tb]
\centering
\includegraphics[width=0.95\textwidth]{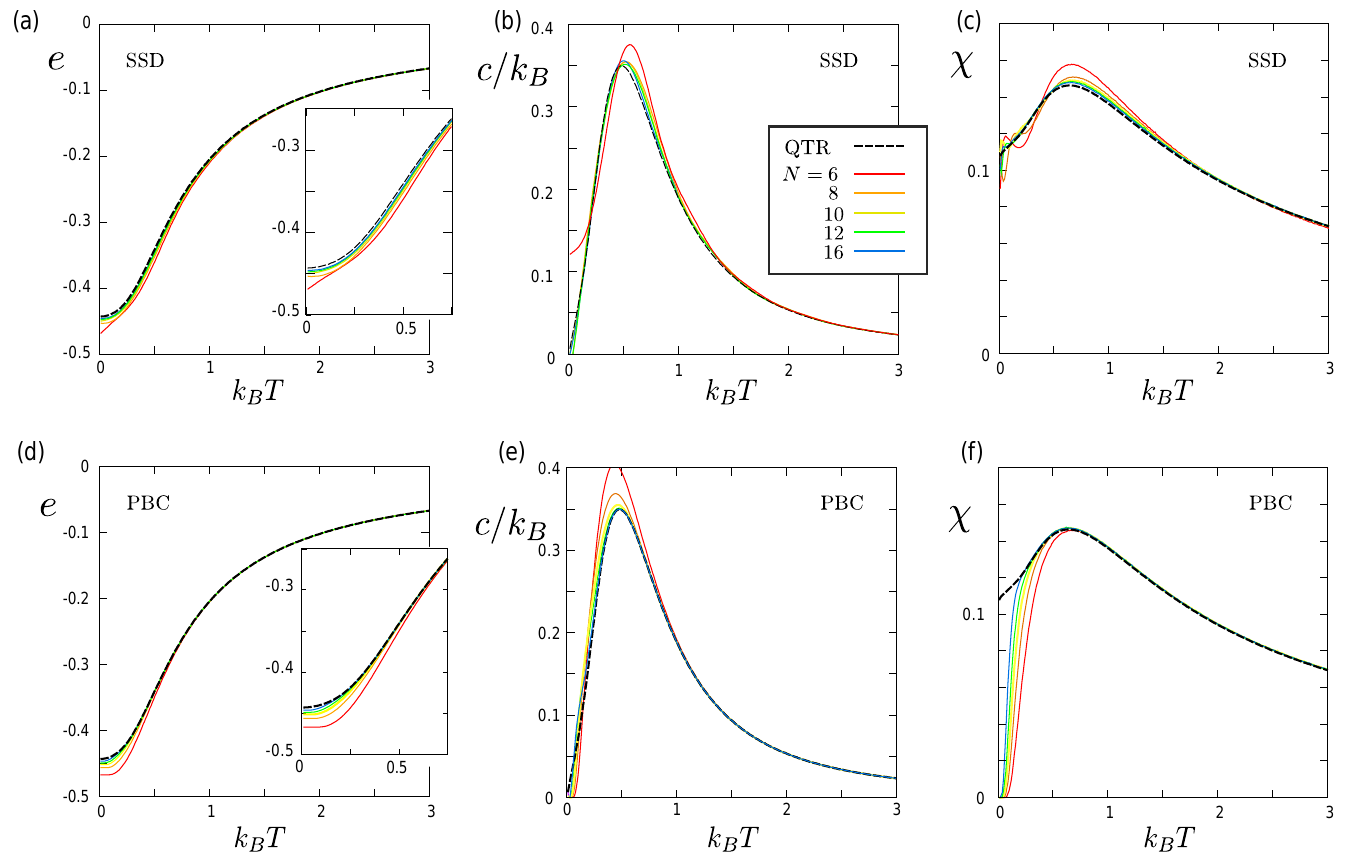}
\caption{
Energy density $e$, specific heat $c/k_B$, and susceptibility $\chi$ of 
the 1D $S=1/2$ AF Heisenberg model for system size $N=6,8,10,12,16$ compared with the 
exact solution using QTM shown in broken line. 
Panels (a)-(c) are obtained using the present SSD framework, 
and (d)-(f) are obtained using the standard ED with PBC. 
}
\label{f2}
\end{figure*}
%*%*%*%*%*%*%*%*%*%%*%*%*%*%*%*%*%*%*%%*%*%*%
%*%*%*%*%*%*%*%*%*%%*%*%*%*%*%*%*%*%*%%*%*%*%

\subsection{1D spin chain}
We first present benchmark results for the 1D antiferromagnetic Heisenberg chain, 
whose exact thermodynamic-limit behavior is precisely known from 
the quantum transfer-matrix (QTM) method 
\cite{klumper1993,klumper1998,klumper2004}. 
Here, we demonstrate the efficiency of our approach by showing 
how rapidly the thermodynamic quantities converge to their $N=\infty$ values 
as we increase the chain length from $N=6$ to $16$, 
which is the range where exact diagonalization (ED) can be reliably applied to the SSD Hamiltonian. 
\par
Figures~\ref{f2}(a--c) compare the SSD results with the exact QTM data. 
For $N=6$, where only $2^6=64$ states are involved, the results deviate 
from the ideal curves below $k_BT \lesssim 1$. 
However, already at $N=8$ (128 states), the SSD results begin to converge 
toward the thermodynamic limit. 
For comparison, using the standard Hamiltonian ${\cal H}$ with PBC, 
clear discrepancies appear below the peaks of $c$ and $\chi$, where both quantities exhibit artificial 
exponential drops caused by the finite-size excitation gap. 
Under PBC, eigenenergies are discretized because the allowed wave numbers satisfy 
$k = 2\pi n / N$, leading to a low-energy excitation gap that scales as $1/N$, 
as is well known from field theory\cite{fisher1972}. 
In contrast, the SSD Hamiltonian suppresses this finite-size gap to scale as $1/N^2$\cite{hotta2013,katsura2012}, 
resulting in a dramatic improvement of finite-size behavior. 
A slight deviation in $c$ and $\chi$ at intermediate temperatures $k_B T \sim 0.5$--$1$ arises because, 
to compensate for the reduced lowest excitation gap, the distribution of energy levels in the 
intermediate energy range becomes somewhat sparse. 
This effect is noticeable for small clusters with $N \lesssim 10$, where fewer than $2^{10}$ 
eigenstates participate, but it rapidly diminishes for larger $N$.

%*%*%*%*%*%*%*%*%*%%*%*%*%*%*%*%*%*%*%%*%*%*%
%*%*%*%*%*%*%*%*%*% fig3 %*%*%*%*%*%*%*%*%*%
\begin{figure*}[tb]
\centering
\includegraphics[width=1.0\textwidth]{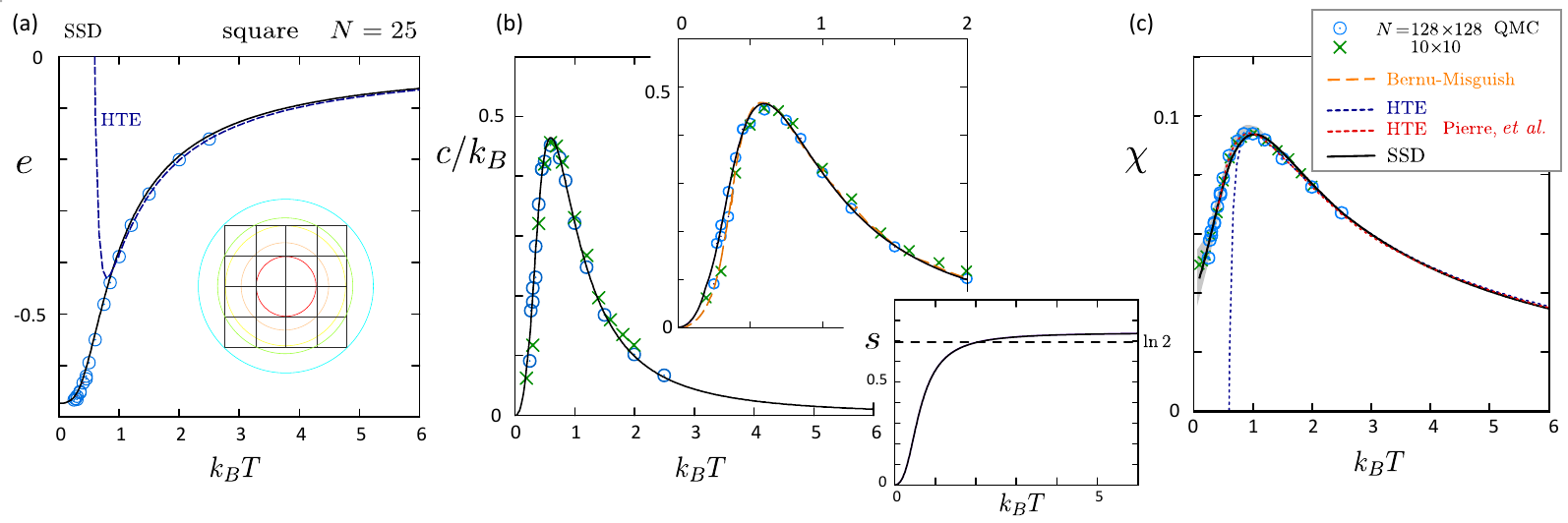}
\caption{
(a) Energy density $e$, (b) specific heat $c/k_B$, and (c) susceptibility $\chi$ of 
the 2D square lattice $S=1/2$ AF Heisenberg model 
based on the SSD analysis with system size $N=5\times 5=25$ given in solid lines. 
The data points are the QMC results using $N=10\times 10$(crosses) and $128\times 128$(circles) clusters, 
and the broken lines are the early HTE\cite{rushbrooks1974}, latest HTE\cite{pierre2024}($\beta^{19}$ for $\chi$), 
and the HTE with interpolation (Bernu-Misguish) \cite{bernu2000}. 
The inset of panel (b) shows the thermal entropy density $s$ obtained by integrating the specific heat. 
}
%% c  Bernu Misguish PRB 63, 134409 (2000)  HTE+interpolation n=13
\label{f3}
\end{figure*}
%*%*%*%*%*%*%*%*%*%%*%*%*%*%*%*%*%*%*%%*%*%*%

%*%*%*%*%*%*%*%*%*%%*%*%*%*%*%*%*%*%*%%*%*%*%
%*%*%*%*%*%*%*%*%*% fig4 %*%*%*%*%*%*%*%*%*%*
\begin{figure*}[tb]
\centering
\includegraphics[width=1.0\textwidth]{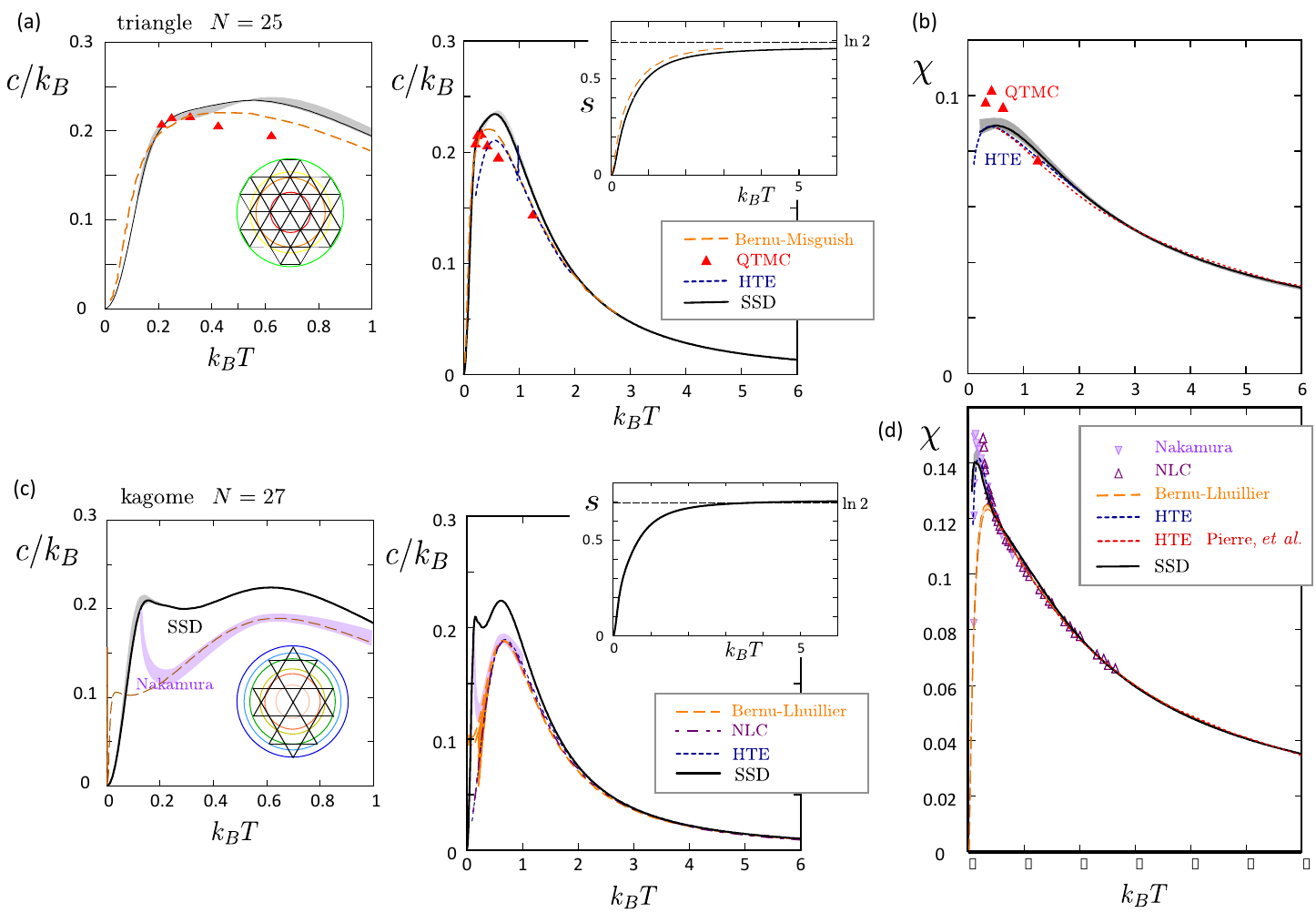}
\caption{
Specific heat $c/k_B$ and magnetic susceptibility $\chi$ of 
the 2D $S=1/2$ AF Heisenberg model on (a,b) triangular and (c,d) kagome lattices shown in solid lines, 
based on the SSD analysis applied to $N=25$ (triangular) and $27$ (kagome) clusters, respectively. 
The references from other methods are shown in several broken lines or data points: 
early HTE \cite{elstner1993}, 
latest HTE\cite{pierre2024}($\beta^{17}$:triangular, $\beta^{19}$:kagome for $\chi$), 
HTE combined with interpolation (Bernu-Misguish) \cite{bernu2000}, 
and QTMC \cite{imada1986} for the triangular lattice, 
and HTE+interpolation (Bernu-Lhuillier)\cite{bernu2015}, 
HTE\cite{lohmann2014}, 
NLC\cite{rigol2006}, 
and QTMC (Nakamura)\cite{nakamura1995} for the kagome lattice. 
For the magnified figure in the range $k_BT\le 0.5$, see Ref.~\onlinecite{hotta2018}. 
%%
%% [c, tri] Bernu Misguish PRB 63, 134409 (2000) n=13   low-t limit: cc=5.32 (Tpaper)^2 = 21.28 kbt^2
%% [cc/chi_lohmann kagome] Lohmann, Schmidt, Richter, PRB 89, 014415 (2014)  HTE16 Pade [7,8]
%% [c, kagome] Rigol, Bryant, Singh, PRL97, 187202 (2006), linked cluster cc
%% [cc/chi_bernu_kagome]  Bernu-Lhuillier, PRL 114, 057201 (2015) 
%%
%% [chi, kagome]  Rigol-Singh, PRL98, 207204 (2007)
%% [tri] Elstner Singh Young PRL71 1629 (1993)  reproduced by code
%% [c, tri]  N. Elstner, R. R. P. Singh, and A. P. Young, PRL71, 1629(1993)
%%
%% [chi, kagome, tri and square]  Pierre et al, SciPost 17, 105 (2024)
}
\label{f4}
\end{figure*}
%
%*%*%*%*%*%*%*%*%*%%*%*%*%*%*%*%*%*%*%%*%*%*%
%*%*%*%*%*%*%*%*%*%%*%*%*%*%*%*%*%*%*%%*%*%*%*%*%*%*%*%*%*%*%*%*%%*%*%*%*%*%*%*%*%*%%*%*%*%
%*%*%*%*%*%*%*%*%*%%*%*%*%*%*%*%*%*%*%%*%*%*%*%*%*%*%*%*%*%*%*%*%%*%*%*%*%*%*%*%*%*%%*%*%*%
\subsection{2D antiferromagnets}
\label{sec:2d-magnets}
%
%*%*%*
We now apply our method to 2D systems. 
Because we employ the full Hilbert space in solving ${\cal H}_{\rm SSD}$, 
the accessible system size is limited to $N \lesssim 30$. 
Since this length scale is extremely small compared with 1D case, 
the reliability of the results requires careful examination. 
To this end, we begin with the square-lattice antiferromagnet, 
for which almost numerically exact reference data based on QMC calculations are available, 
and after that apply it to the frustrated lattices.  

\subsubsection{square lattice Heisenberg antiferromagnet}
Figures~\ref{f3}(a)-\ref{f3}(c) show the temperature dependence of the energy density $e$, 
specific heat $c/k_B$, and magnetic susceptibility $\chi$ (solid lines), 
obtained from the SSD Hamiltonian on a $N=5\times 5$ square lattice cluster. 
For comparison, we also plot QMC data for system sizes $128\times 128$ \cite{wang1992} and $10\times 10$ \cite{okabe1988}, 
together with the HTE results \cite{rushbrooks1974,pierre2024} (broken lines). 
Our SSD data agree with the large-scale QMC results within $10^{-3}$.
\par
For $k_BT \gtrsim 1$, the HTE serves as a useful reference to assess the quality of our results; 
the one obtained in the early stage remains accurate 
before deviating from the true curve at $k_B T \lesssim 1$\cite{rushbrooks1974}. 
Recent update of HTE on $\chi$ using higher orders and coinciding Pad\'e approximants 
reaches down to $k_BT\sim 0.5$\cite{pierre2024}. 
To obtain reliable low-temperature specific heat, Bernu and Misguich proposed combining HTE ($\beta^{13}$) with 
sum rules for entropy, $\int c\,dT/T = \ln 2$, and for energy, $\int c\, dT = e(T=\infty)-e(0)$, 
and interpolated the data smoothly from high-temperature to the ground state. 
We denote this as HTE+interpolation. 
Their results (broken line) agree well with QMC down to $k_BT \sim 0.2$.

In our SSD calculation, evaluating $e$ for $k_BT \lesssim 0.4$ is challenging: 
numerical inaccuracies come from the choice of the solver, 
and leads to a slight overestimate of the specific heat 
relative to the HTE+interpolation results, and consequently to a larger total entropy, 
$\int_0^6 c\, dT/T = 1.05 \ln 2$. 
Nevertheless, the overall thermodynamic behavior is nicely reproduced. 
%% ln2=0.69314718056
\par

\subsubsection{frustrated antiferromagnets}
We now apply the method to the frustrated triangular and kagome lattices, 
using circular clusters of size $N=25$ and $27$, respectively.
In Fig.~\ref{f4} we plot the specific heat and magnetic susceptibility over a temperature range of $k_B T =0$ to 6. 
For comparison, we include several reference data sets obtained by different numerical methods: 
the HTE with Pad\'e approximation\cite{elstner1993,lohmann2014,pierre2024}, HTE+interpolation\cite{bernu2000,bernu2015}, 
and the NLC approach\cite{rigol2006}.
The NLC method systematically incorporates clusters of up to $n=8$ triangles (corresponding to $2n+1$ sites) 
and generally has an advantage over HTE of the same order in the lower-temperature regime. 
Indeed, for both kagome and triangular lattices, the earlier HTE results\cite{elstner1993,lohmann2014} 
start to deviate from the reliable curves at $k_BT \lesssim 1$, 
where the data becomes sensitive to the expansion order and the Pad\'e approximation\cite{lohmann2014}. 
The latest HTE with $\beta^{17}-\beta^{19}$ and coinciding Pad\'e approximants 
sustains down to $k_BT \sim 0.5$\cite{pierre2024} for both lattices.
The NLC data further reaches $k_BT\sim 0.1$ 
consistently with other data down to $k_BT \sim 0.5$\cite{rigol2006,rigol2007}. 
The quantum transfer-matrix Monte Carlo (QTMC) method\cite{imada1986} 
evaluates finite-temperature quantities via imaginary-time evolution with basis states 
generated by Monte Carlo sampling.
For the triangular lattice, QTMC calculations were performed up to sizes $N=3 \times 6$ and 
$2 \times 8$ and then extrapolated to $N\rightarrow\infty$\cite{imada1987}. 
For the kagome lattice, QTMC simulations have been carried out up to $N=72$\cite{nakamura1995}, 
where the sign problem was avoided by tracing out part of the sublattices.
\par
As our numerical solver does not provide sufficiently accurate results at the lowest temperatures, 
particularly for $T \le 0.1$, we use the data for $T \ge 0.2$, 
and assume that the specific heat follows the form $c \propto T^{2}$ at the ground state. 
We apply a kernel regression method\cite{harada2011,nakamura2016} 
to extrapolate the data down to zero temperature
%%%%%%%%%
\footnote{
For low temperature calculation, often the low-energy spectra obtained by ED is applied\cite{morita2020}. 
However, the cluster size available for ED is small, and even with SSD, the oscillations induced from the 
system edges penetrates throughout the system, which are difficult to remove without bias, particularly at $T\lesssim 0.1$. 
The kernel regression gives a more reasonable results, as far as we assume the smooth temperature dependence 
of the data. 
}.
%%%%%%%%%
After this extrapolation, the quantity $c/T$ exhibits a peak at $T \simeq 0.16$ and $0.1$ 
for the triangular and kagome lattices, respectively. 
In both lattices, the specific heat shows extended peaks 
(or a merged double-peak structure), while the kagome lattice has a more visible double-peak.
Indeed, such peak structures are unavoidable if one aims to obtain a reasonable extrapolation for $k_BT \le 0.2$. 
The entropy density integrated over the plotted range becomes
$\int_0^6 S dT = 0.948 \ln 2$ for the triangular lattice and $1.02 \ln 2$ for the kagome lattice.
Accordingly, the kagome lattice should exhibit a smaller specific heat at $k_BT \lesssim 0.1$ than what we currently obtain, 
likely closer to the QTMC results of Nakamura {\it et al.}\cite{nakamura1995}.
Nevertheless, the overall peak positions appear to capture the intrinsic thermodynamic features expected in the bulk systems.
\par 
There are several literatures reporting the specific heat of the triangular lattice antiferromagnet. 
In the Heisenberg antiferromagnet, the XTRG by Chen, {\it et al.} shows the extended peak 
in the range $T_{\rm peak}=[0.2:0.55]$\cite{chen2019}. 
These peak positions are similar to our results, $T_{\rm peak}=[0.2:0.6]$ in Fig.~\ref{f4}(a). 
They ascribe these extended peak region to the roton-like excitation 
that accompanies chirality and substantial density of states, 
which is actually observed with a gap of $\sim 0.55J$ at the $M$ point in the excitation spectrum \cite{weihong2006}. 
A similar extended peak structure is observed in materials like 
Na$_2$BaCo(PO$_4$)$_2$\cite{li2020,gao2022} 
and Ba$_8$CoNb$_6$O$_{24}$\cite{rawl2017,cui2018} with different XXZ exchange anisotropies 
both in theory and in experiment. 
These results show that the extended-peak structure is robust against the change of the model parameters, 
intrinsic to the triangular lattice antiferromagnet. 
The features of the long-range 120$^\circ$ order or supersolid order of the Heisenberg and XXZ antiferromagnets 
appear well below this peak temperature range. 
\par
For the kagome lattice, 
previously, the HTE+interpolation assuming a gapped behavior 
(Bernu--Lhuillier in Fig.~\ref{f4}(c)\cite{bernu2015})
predicted double peaks with
$(T_{\rm peak},c_{\rm peak})=(0.01,0.1)$ and $(0.5,0.18)$.
More recently, the lower-temperature peak has been updated under the assumption of a
gapless ground state, giving
$(T_{\rm peak},c_{\rm peak})=(0.05\text{--}0.1,0.1)$\cite{messio2020}.
The FTLM \cite{schnack2018} up to $N=42$ sites also reports a 
shoulder-like double-peak structure around $(T_{\rm peak},c_{\rm peak})=(0.06-0.7,0.1-0.2)$. 
They also show a series of small structures at lower temperature, which have cluster-dependence 
and may not be intrinsic. 
\par
In our SSD case, the double peaks appear at 
$(T_{\rm peak},c_{\rm peak})\sim (0.1,0.2)$ and $(0.6\text{--}0.7,0.2)$, 
whose positions are close to the QTMC results\cite{nakamura1995}. 
Since our analysis assumes neither a gapped nor a gapless ground state, and relies only on 
a natural extrapolation of $c/T$ data toward $T=0$ with its leading order linear in $T$, 
it is essential to obtain a double-peak structure when enforcing a low-temperature extrapolation.
Therefore, it is reasonable to conclude that our results exhibit distinct and robust
double peaks at $T_{\rm peak}\sim 0.1$ and $0.6-0.7$. 
\par
There has been a debate on whether the kagome Heisenberg antiferromagnet
possesses a double (or multiple) peak structure. 
In simple ED or FTLM calculations, the low-energy structures are strongly dictated by
the cluster shape available for system sizes $N\sim 27$--$41$. 
Consequently, two or more small peaks typically appear below the major peak at
$T_{\rm peak}\sim 0.8$\cite{sugiura2013,shimokawa2016,morita2020,schnack2018}. 
Because systematic size scaling is extremely difficult in these methods, it has been
difficult to determine whether such peaks survive at $N=\infty$, 
while we think it is likely to be a size effect. 
\par
The susceptibility $\chi$ continues to increase down to the temperature range 
$k_BT \lesssim 0.5$ for both lattices, which is characteristic of frustrated magnetism. 
The peak value is significantly larger for the kagome lattice than for the triangular lattice. 
For the triangular lattice susceptibility, only the HTE results\cite{elstner1993} and a small
number of extrapolated QTM data points for clusters up to $N=4 \times 4$\cite{imada1987} 
were available, leaving the profile of $\chi$ for $k_{B}T\lesssim 1$ unclear. 
Consequently, comparisons with experimental results often
relied on HTE analyses\cite{weihong2005,yoshida2015}. 
As seen in Fig.~\ref{f4}(b), the HTE and QTM results exhibit noticeable discrepancies for 
$k_{B}T \lesssim J$, whereas our SSD-based susceptibility lies between these two curves.

%*%*%*%*%*%*%*%*%*%*%*%*%*%*%*%*%*%*%*%*%*%*%*%*%*%*
\section{Summary}
We have obtained the temperature dependence of the specific heat and magnetic susceptibility
of the quantum $S=1/2$ Heisenberg models on 1D, square, and frustrated triangular and kagome lattices,
using the SSD analysis we proposed~\cite{hotta2012,hotta2013,hotta2018} 
in combination with the TPQ approach.
Because the specific heat at $k_BT \lesssim J$ is typically very sensitive to finite-size effects
and to the choice of numerical solver, and is also not directly accessible by HTE,
there have been only a few reports particularly for the triangular lattice. 
We confirmed a good agreement of our results using $N=25$ cluster with QMC with $N=128\times 128$ on a square lattice. 
For both kagome and triangular lattices, the specific heat exhibits an extended peak structure
in the range $k_B T/J \sim 0.1$--$0.7$, with the kagome lattice showing a more pronounced
double-peak structure. 
Although the peak heights are not necessarily consistent among different methods, 
the presence of two peaks and their locations around $T_{\rm peak}/J \sim 0.1-0.2$ and $0.7$
should be regarded as well established, as they are common to QMC, HTE with interpolation,
and our SSD results, as well as XTRG ones. 
The magnetic susceptibility also displays characteristic features of frustrated magnetism:
it increases gradually toward low temperatures and reaches its maximum at
$T_{\rm peak} \sim 0.2$--$0.5$.
\par
Both kagome and triangular lattices retain large entropy down to very low temperatures, 
reminiscent of their classical Ising counterparts with residual entropy~\cite{wannier1950,kano1953}. 
However, the nature of this entropy appears to differ: 
the triangular lattice likely contains a larger fraction of nonmagnetic states in its low-energy 
manifold within $\Delta E \lesssim 0.5J$, 
whereas the kagome lattice hosts strong competition among numerous magnetic states. 
Such a qualitative difference in the origin of the entropy may naturally lead to distinct 
ground-state properties and excitation spectra in these two representative frustrated systems. 
In a series of XXZ or Heisenberg triangular lattice antriferromagnets, 
there are gapped rotons with large density of states\cite{chen2019,li2020,gao2022}, 
and the magnetism either being N\'eel order or supersolid, is determined below that energy scale. 
The kagome lattice antiferromagnet can be gapless with a structureless 
excitation spectrum\cite{endo2018,punk2014} reminscent of spinons, indicating a more serious competiton 
among different magnetic configurations.

%*%*%*%*%*%*%*%*%*%*%*%*%*%*%*%*%*%*%*%*%*%*%*%*%*%*
\begin{acknowledgments}
I would like to express my sincere gratitude to Johannes Richter 
for the heartwarming communications and stimulating discussions during the past years, 
which I miss a lot. 
The present work started at around 2016, and remained unpublished 
after the first publication in 2018 coherently with Johannes's work, 
and this special issue has pushed me forward to finally write it up. 
It was supported by a Grant-in-Aid for Transformative Research Areas 
``The Natural Laws of Extreme Universe---A New Paradigm for Spacetime and Matter from Quantum Information" 
(Grant No. 21H05191) and other JSPS KAKENHI (No. 21K03440). 
The numerical simulations were performed 
on the Yukawa-21 supercomputer at the Yukawa Institute for Theoretical Physics, Kyoto University, 
and at the Supercomputer Center, the Institute for Solid State Physics, 
the University of Tokyo.  
\end{acknowledgments}

%%%%%%%%%%%%%%%%%%%%%%%%%%%%%%%%%%%%%%%%%%%%
%%%%%%%%%%%%%%% bibliography %%%%%%%%%%%%%%%
%%%%%%%%%%%%%%%%%%%%%%%%%%%%%%%%%%%%%%%%%%%%
\bibliography{ftcalc}
\end{document}